\pdfoutput=1
\documentclass[twocolumn,aps,prb,superscriptaddresses]{revtex4-1}
\usepackage{graphicx}
\usepackage{amsmath,bm}
\usepackage{comment}
\usepackage[colorlinks]{hyperref}
\usepackage{bbold}
\usepackage{xcolor}
\usepackage{natbib}

\hypersetup{
	colorlinks,
	citecolor=blue,
	filecolor=blue,
	linkcolor=blue,
	urlcolor=blue}

\begin{document}	
	\title{Entangling Spins in Double Quantum Dots and Majorana Bound States}
	\author{Marko J. Ran\v{c}i\'{c}$^{1}$}
	\author{Silas Hoffman$^{1}$}	
	\author{Constantin Schrade$^{2}$}
	\author{Jelena Klinovaja$^{1}$}
	\author{Daniel Loss$^{1}$}
	\address{$^{1}$Department of Physics, University of Basel, Klingelbergstrasse 82, 4056, Basel, Switzerland}	
   	\address{$^{2}$Department of Physics, Massachusetts Institute of Technology, 77 Massachusetts Ave., Cambridge, MA 02139}
	\date{\today}
	
	\begin{abstract}
We study the coupling between a  singlet-triplet qubit realized in a double quantum dot to a topological qubit realized by spatially well-separated Majorana bound states. We demonstrate that the singlet-triplet qubit can be leveraged for readout of the topological qubit and for supplementing the gate operations that cannot be performed by braiding of Majorana bound states. Furthermore, we extend our setup to a network of singlet-triplet and topological hybrid qubits that paves the way to scalable fault-tolerant quantum computing. 
	\end{abstract}
	\maketitle

\section{Introduction}
	
Quantum dots (QDs) are promising scalable candidates for realizing quantum bits. By using spin degrees of freedom to store quantum information, experiments have realized fast, universal control of, so-called spin qubits \cite{Loss1,Petta1,Hanson1,Reilly1, Bluhm2, Bluhm1,Kawakami1,Muhonen2014storing,Scarlino1}. The same couplings that enable the ease and speed in control of the spin qubit are simultaneously responsible for unwanted incoherent noise and subsequent decoherence of the qubit. Despite improvements in decoherence  times, several decoherence mechanisms remain for spin qubits\cite{Bluhm1,Bluhm2,Kloeffel2013prospects,Reed1} that causes operational errors and necessitates correction and ultimately decreases the operational speed of a quantum computer.

In particular, single electron qubits use the collinear spin states of a single electron level, split by the Zeeman effect, to furnish a qubit and time-dependent magnetic or electric fields to realize qubit rotations \cite{Loss1,Kloeffel2013prospects}. Consequently, fluctuations in the local magnetic fields, due to nuclear magnetic moments, presence of magnetic impurities, and variations in gate voltages decrease decoherence  times \cite{Merkulov1}. Alternatively, two electron levels, often furnished by a double quantum dot (DQD), can be used to store information in the singlet-triplet ($ST_0$) subspace\cite{Levy1}. Rotations about orthogonal axes of the  Bloch sphere are performed by manipulating the interdot exchange interaction and inducing a difference in the Zeeman splitting between the orbital levels, achieved by embedding a micromagnet on top of the DQD or by $g$-factor engineering \cite{Tokura1,Pioro1,Wu1}. Although this means that these qubits are susceptible to both electric and magnetic noise, the former can be reduced by operating the qubit ``symmetrically'' \cite{Martins1, Reed1} wherein the charge noise is minimized. As a result, current state-of-the-art semiconductor $ST_0$ qubits have a decoherence time of $T_2^*=1\text{ $\mu$s}$\cite{Martins1, Reed1}, as compared to single electron spin qubits which have a decoherence time of $T_2^*=10\text{ ns}$\cite{Merkulov1}. 

\begin{figure}[t!]
	\includegraphics[width=0.4\textwidth]{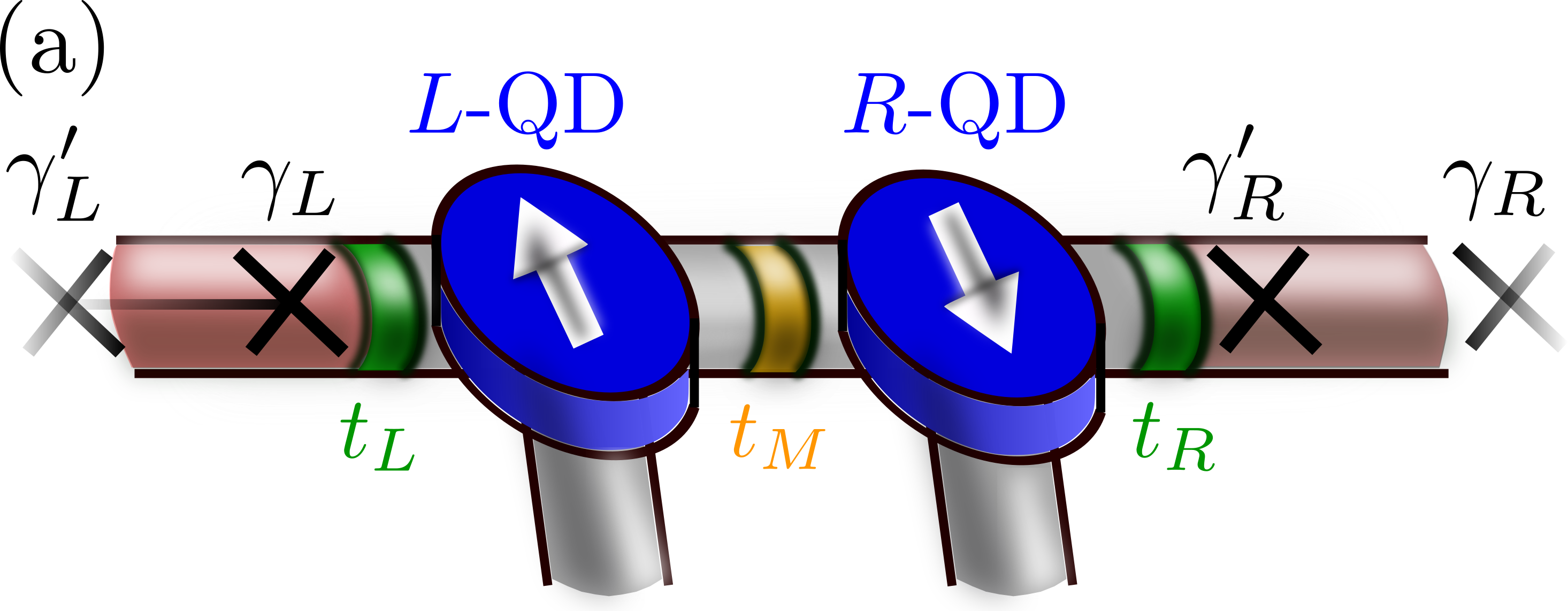}
	\includegraphics[width=0.4\textwidth]{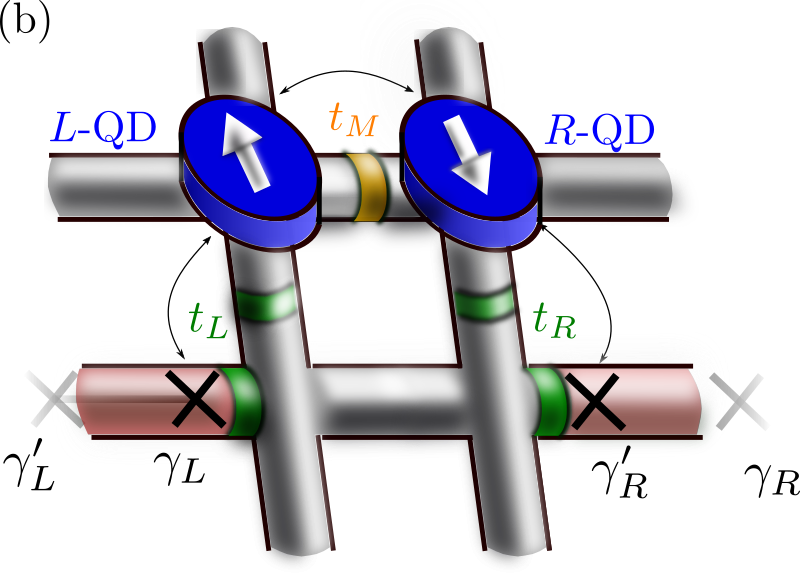}
	\caption{(Color online) Two possible physical realizations of our proposal. (a) The linear arrangement with two quantum dots (blue cylinders), $L$-QD and $R$-QD, and two TNWs (red) hosting MBSs at their ends, $\gamma_L$ and $\gamma_R'$, which are realized in the same NW. There are two additional NWs that allow for braiding of the MBSs. The amplitude of the coupling between $\gamma_L$ ($\gamma_R'$) and $L$-QD ($R$-QD), $t_L$ ($t_R$), is controlled by a local backgate (green). The interdot tunneling $t_M$ is similarly controlled by a backgate (orange). (b) The same model can be realized by four NWs forming a hashtag. In this case, the QDs are realized at the intersection of the two vertical NWs with the upper horizontal NW. The TNWs are on the left and right sides of  the lower wire (red). We assume a large controllable barrier between the left and right TNWs so there is no overlap between $\gamma_L$ and $\gamma_R'$. The physical parameters are defined analogously to the first setup. We note that the ``hashtag" geometry requires two tunnel barriers to control the coupling between the dots and the MBS.}\label{Fig1}
\end{figure}

Departing from spin qubits entirely, there have been several proposals to realize Majorana bound states (MBSs)\cite{PhysRevLett.100.096407,RevModPhys.80.1083,Oreg1,Sato2,PhysRevLett.105.077001,PhysRevB.88.020407,PhysRevLett.111.147202,PhysRevLett.111.186805,PhysRevLett.111.206802,PhysRevB.88.155420,Sato1} and to utilize their occupancy as the basis for a qubit \cite{Kitaev2,Hassler2}. The MBSs are charge- and spinless bound states at zero energy that obey non-Abelian braiding statistics and are believed to be robust to local perturbations  {\color{blue} \cite{Kitaev1,PhysRevB.86.085408,PhysRevB.85.235307,PhysRevB.86.140503,PhysRevLett.108.257001,PhysRevLett.108.260504,PhysRevLett.109.236801,PhysRevB.86.180503,PhysRevB.88.165111,PhysRevB.88.155133,PhysRevB.90.195421,PhysRevLett.115.237001,PhysRevB.93.165116,PhysRevB.96.085418,escribano2017interaction}}. As a result, in contrast to spin qubits, MBS qubits are largely immune to decoherence effects arising from fluctuations in the electromagnetic fields or due to quasiparticle poisoning\cite{PhysRevB.84.205109,PhysRevB.86.085414,PhysRevB.85.174533,PhysRevB.85.174533,PhysRevB.85.121405,PhysRevB.97.125404,Aseev1,PhysRevB.97.054508}.

Eigenstates of the MBS qubit are degenerate, making energy-resolved readout of the MBS qubit impossible. Furthermore, decoherence-free operation of MBS qubits, achieved by exchanging pairs of MBSs, is limited to a non-universal set of quantum gates \cite{Bravyi1}, which can be efficiently simulated with a classical computer \cite{gottesman1}. This is why the investigation of setups where QDs serve as a toolkit to control MBS qubits and provide information about MBSs has been an active field of research\cite{Sau1,Hassler1,PhysRevB.84.140501,PhysRevB.84.201308,Pekker1,PhysRevLett.111.036802,PhysRevB.89.245413,PhysRevB.89.165314, You1, Silas1, Plugge1,Landau1,deng2016majorana,szombati2016josephson,Plugge2,ptokPRB17,PhysRevB.97.045404,dengPRB18}.

In this work, we propose tunnel-coupling two topological nanowires (TNWs), hosting a MBS qubit, with a DQD loaded with two electrons (see Fig.~\ref{Fig1}). Upon projecting the system to a low-energy subspace, consisting of two two-level systems, the MBS qubit and the $ST_0$ qubit, we find an effective Hamiltonian capable of entangling the two qubits. After detailing the operations to realize the gates necessary for universal quantum computation, we are able to calculate the fidelity of those gates using parameters that are consistent with $ST_0$ and MBSs qubits in nanowires \cite{Kitaev1, Mourik1}. Although our discussion focuses on MBSs and spin qubits formed in nanowires (NWs), our findings are more general and apply for different realizations of MBS qubits and spin qubits, \textit{e.g.}, a chain of magnetic adatoms \cite{Nadj2014observation, Pawlak1, Ruby1} coupled to a $ST_0$ qubit in a phosphorus donor \cite{kane1998silicon,PhysRevB.82.241302,pakkiam2018characterisation}.

Before discussing the details of our setup, we wish to highlight several improvements upon previous proposal in which single electron spin QDs couple to MBSs\cite{Silas1}. (1) The $ST_0$ qubit couples only to magnetic field gradients. This avoids unintentional couplings to the global magnetic field which is used to generate the MBSs. (2) The $ST_0$ qubits can be readout by non-invasive charge measurements without exchanging  electrons with the leads, allowing much faster initialization of the $ST_0$ qubit compared to the single electron spin qubit \cite{petta2005coherent,PhysRevLett.110.046805}.

This paper is organized as follows.~In Sec.~\ref{sec:Setup} we describe our setup and the corresponding Hamiltonian which, in Sec.~\ref{sec:LowEn}, we reduce to a low-energy effective Hamiltonian acting only in the space of the qubits. In Sec.~\ref{sec:Universality} we discuss the quantum gates resulting from the coupling between the qubits, single qubit operation and subsequently quantify their fidelity in Sec.~\ref{sec:Decoherence}. Our scheme is generalized to a network of coupled $ST_0$-MBS qubits in Sec.~\ref{sec:Network}, before concluding in Sec.~\ref{sec:conclusion}.

\section{Setup}\label{sec:Setup}
Our setup is comprised of two QDs, forming a DQD, coupled to two topological superconductors, both of which are realized within NWs. The NWs are coupled to a bulk superconductor and subjected to a global magnetic field capable of driving the NWs into the topological phase. Furthermore, we assume that there are a series of gates along the NWs that can (1) selectively define the segments of the NW in the topological phase which we henceforth refer to as topological nanowire segments (TNWs), (2) control energy levels of the quantum dots, and (3) mediate the tunneling of electrons between dots and TNWs. 

We consider two possible physical realizations of such a system, both of which are well-described by the Hamiltonian detailed in this section and therefore amenable to analyses in the remainder of the paper. In the first setup, which we refer to as an linear arrangement, a DQD is gate-defined between two TNWs within the same NW [see Fig.~\ref{Fig1}(a)].\cite{su2017andreev,saldana2018supercurrent} Two auxiliary quantum wires are attached, forming a double T-junction, which allow for braiding of the MBSs. In the second setup, four NWs form a so-called hashtag structure\cite{gazibegovic2017epitaxy} [see Fig. \ref{Fig1} (b)]. In this arrangement, the quantum dots are located at the intersections of the upper horizontal wire with the two parallel vertical wires, while the left and right segments of the lower horizontal nanowire are tuned into the topological regime (depicted in red).

We consider QDs that are sufficiently small so that only a single orbital level of the QD is relevant. The Hamiltonian of the DQD is 

\begin{align}\label{eq:HamDD}
H_{\text{\rm DQD}}&=\sum_{j,s} \left[\left(\epsilon_{0}+sE^{\text{Z}}_{j}\right)d_{j,s}^\dagger d_{j,s}+Vn_{j,s}n_{j,\bar{s}}/2\right]\\
&+\sum_{s}\left[t_{M,s}d^{\dagger}_{L,s}d_{R,s}+\text{H.c.}\right]+\sum_{s,s'} \tilde V n_{R,s}n_{L,s'}\nonumber,
\end{align}
where $d_{j,s}$ ($d_{j,s}^\dagger$) annihilates (creates) an electron with spin $s$, quantized along the magnetic field, on QD $j$ ($j$-QD). Here, $j=L (R)$ refers to the left (right) QD and $s=\pm1/2$ for up and down spin. The level position and intradot charging energy are denoted by $\epsilon_0$ and $V$, respectively, where $n_{j,s}=d_{j,s}^{\dagger}d_{j,s}$.  The singly occupied charging energy is denoted by $\tilde{V}$. The Zeeman splitting, $E_j^\textrm{Z}$ on $j$-QD, must necessarily be different to operate the $ST_0$, as we see below, and can be tuned by applying local magnetic fields\cite{Tokura1,Pioro1} or by tuning of the $g$-factor\cite{trif2,Salis1,Schroer1,Kloeffel2013prospects,olesia1}.
Upon making an appropriate gauge choice, the interdot tunneling, $t_M$, is real without loss of generality.
 
When the length of the TNW is comparable to the decay length of the MBS, the MBSs overlap and their coupling is characterized by a phenomenological parameter $\Xi_j$. Furthermore, we operate in the regime in which the coupling between the DQD and supergap states can be neglected (the conditions under which this is valid are discussed in detail in Sec.~\ref{sec:LowEn}) such that the Hamiltonian describing the TNW simplifies to $H_{\rm TNW}=\sum_{j}i\Xi_j \gamma_j\gamma_j'$, where $j=L(R)$ stand for left(right) TNW in the context of Fig. \ref{Fig1}.

The MBSs in the TNWs weakly overlap with the electronic states of the QDs, described by the following tunneling Hamiltonian,
\begin{equation}\label{eq:HamT}
	H_{\text{T}}=\sum_{s}\left(t_{L}d^{\dagger}_{L,s}\gamma_{L}+t_{R}d^{\dagger}_{R,s}\gamma_{R}'+\text{H.c.}\right),
	\end{equation}
where $t_L$ ($t_R$) is the matrix element for an electron on $L$-QD ($R$-QD) to tunnel into the rightmost (leftmost) MBS, $\gamma_L$ ($\gamma_R'$), in the left (right) TNW. Again, the tunneling amplitude $t_{j}$ is taken to be real, according to an appropriate gauge choice, assumed to be spin-independent \cite{silas12}. The total Hamiltonian is given by $H=H_{\rm TNW}+H_{\text{DQD}}+H_{\rm T}$.
	
\section{The Effective Qubit Hamiltonian}\label{sec:LowEn}
We operate the quantum dot in the regime ${|E^{\text{Z}}_{L}-E^{\text{Z}}_{R}|\ll t_M \ll V, E_j^{\rm Z}}$, where the two levels in the DQD closest to the chemical potential of the TNW are the spin singlet state and the spinless spin triplet state,
\begin{align}
|S\rangle=\frac{1}{\sqrt{2}}\left(d_{L,\uparrow}^\dagger d_{R,\downarrow}^\dagger-d_{R,\uparrow}^\dagger d_{L,\downarrow}^\dagger\right)|0_{\rm DQD}\rangle\,,\nonumber\\
|T_0\rangle=\frac{1}{\sqrt{2}}\left(d_{L,\uparrow}^\dagger d_{R,\downarrow}^\dagger+d_{R,\uparrow}^\dagger d_{L,\downarrow}^\dagger\right)|0_{\rm DQD}\rangle\,,
\end{align} 
respectively, with $|0_{\rm DQD}\rangle$ denoting the vacuum state of the DQD, and that all other states in the quantum dot are sufficiently separated.~These states provide a low-energy subspace of the DQD which furnishes a two-level system for the spin qubit. 
	
Each pair of MBSs in the TNWs forms a nonlocal complex fermion, $f_{j}=(\gamma_{j}+i\gamma'_{j})/2$. For a TNW that is longer than the localization length of the MBS which we henceforth assume, \textit{i.e.} $\Xi_j=0$ \cite{PhysRevLett.108.257001,PhysRevLett.111.056802,diego}, this fermion lies at zero energy so that the occupancy and vacancy of this level is degenerate. As the computational basis of the MBS qubit must be of fixed parity, we choose the odd total parity sector in which the nonlocal complex fermion in either the left or right TNW must be occupied.  The two levels of the MBS qubit are $|L\rangle=f^\dagger_L|0_{\rm TNW}\rangle$ and $|R\rangle=f^\dagger_R|0_{\rm TNW}\rangle$, with $|0_{\rm TNW}\rangle$ being the vacuum state of the TNW. We assume that the $ST_0$ qubit is coherently coupled to the MBS qubit and decoupled from states above the gap $|\epsilon_0\pm E_j^{\rm Z}/2|\ll\Delta$, with $\Delta$ being the proximity induced gap of the topological superconductor, which should be close to the superconducting gap of the parent $s$-wave superconductor in the strong coupling regime \cite{sc1, sc2,reegSC1,reegSC2,sc3,sc4,sc5}.
 Also we assume that the DQD has a well defined number of confined charge carriers - that it is protected from leakage to states with a different number of electrons in the DQD, $t_j/|\epsilon_0\pm E_j^{\rm Z}/2|\ll 1$. The total low energy subspace of our system, which furnishes our full computational basis, consists of the following four states: $\{|S\rangle |L\rangle,\,|S\rangle |R\rangle,\,|T_0\rangle |L\rangle,\,|T_0\rangle |R\rangle\}$.

 \begin{figure}[b!]
 	\includegraphics[width=0.45\textwidth]{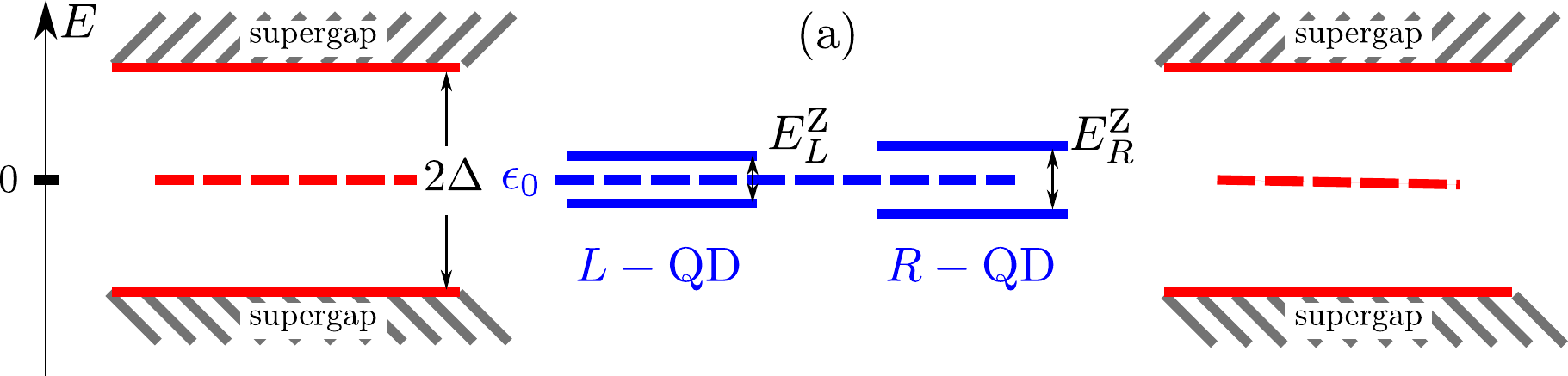}\vspace{0.5 cm}
 	\includegraphics[width=0.45\textwidth]{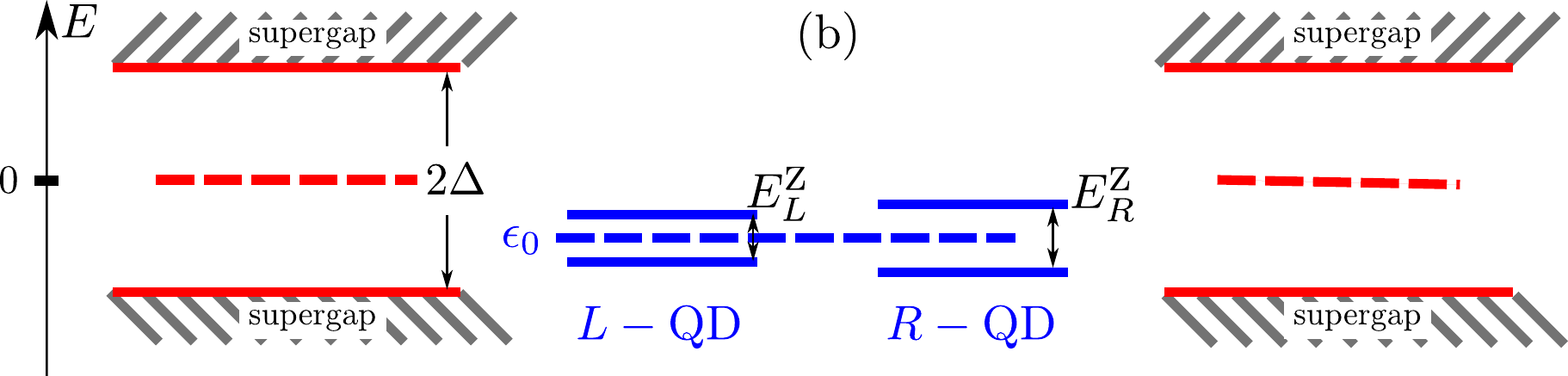}
 	\caption{ Schematic energy diagram demonstrating two possible regimes of the coupling between QDs and MBSs. (a) In the first regime, the single particle levels of the QDs are tuned close to MBS energies,  ${\epsilon_{0},\tilde{V}} \ll E_{R,L}^Z/2\ll \Delta$.
(b) In the second regime, the single particle levels of QDs are tuned away from zero,  $E_{R,L}^Z \ll {\epsilon_{0},\tilde{V}} \ll \Delta$. Here, $\Delta$ is the proximity induced gap, $\epsilon_0$ is the detuning of the DQD levels with respect to the zero of energy  of MBSs, and $E_j^{\rm Z}$ the Zeeman splittings on the QDs.}\label{Fig:EnDiag}
 \end{figure}

Within the limits of the parameters given at the beginning of this section, we perform a Schrieffer-Wolff transformation \cite{Winkler1,bravyi2011schrieffer}, retaining terms up to cubic order in the tunnelings, first order in difference in Zeeman splitting between the two dots and neglecting all terms which scale as a product of two tunneling hoppings and difference of the DQD Zeeman splitting. As a result, we obtain an effective Hamiltonian
\begin{equation}\label{eq:LowEH}
\mathcal{H}=J\sigma_z+\delta \sigma_x+D\sigma_y\eta_y\,,
\end{equation}
where $\sigma_i$ ($\eta_i$) are the Pauli matrices acting on the  $ST_0$ (MBS) qubit subspace while leaving the MBS ($ST_0$) qubit subspace unchanged, \textit{i.e.}, $|S\rangle$ and $|T_0\rangle$ ($|L\rangle$ and $|R\rangle$) are eigenvectors of $\sigma_z$ ($\eta_z$). In the limit when ${|E_L^{\rm Z}-E_R^{\rm Z}|\ll V-\tilde{V}}$ (with $V>\tilde V$),

\begin{equation}\label{eq:J}
J=\frac{2 t_M^2}{V-\tilde{V}}\,,
\end{equation}
 is half of the singlet-triplet splitting arising from the exchange interaction between the two QDs~\cite{Loss1}. As $J$ is independent of the coupling between QDs and MBSs, this exchange interaction occurs independent of the TNWs and is used to control rotations about the $z$ axis of the $ST_0$ Bloch sphere. The second term has a coefficient 

 \begin{equation}\label{eq:delta}
 \delta=
 \frac{E^{\rm Z}_L-E^{\rm Z}_R}{2}\,,
 \end{equation}
 where we keep only the leading term given by the asymmetry in the Zeeman splitting between the two QDs and neglect high-order corrections. The last term in $\mathcal{H}$ [see Eq.~(\ref{eq:LowEH})] is an entangling coupling between the $ST_0$ and MBS qubit with
\begin{equation}\label{eq:DwithV}
D=
\frac{t_Lt_Rt_M\left(E_L^{\rm Z}+E_R^{\rm Z}\right)\left(\frac{E_L^{\rm Z}E_R^{\rm Z}}{4}+\tilde{V}v_2+\epsilon_{0}(v_3-\epsilon_0)\right)}
{v_1
\left(\frac{\left(E_L^\textrm{Z}\right)^2}{4}-(\tilde{V}+\epsilon_0)^2\right)
\left(\frac{\left(E_R^\textrm{Z}\right)^2}{4}-(\tilde{V}+\epsilon_0)^2\right)}
\,,
\end{equation} where $v_n=V-n\tilde{V}$ with $n=1,2,3$.

This coupling is the result of two types of processes. In the first process, (1) an electron in the $L$-QD tunnels into the left TNW, (2) an electron from the right TNW tunnels into the $R$-QD and 
(3) one of the electrons in the doubly occupied $R$-QD tunnels into the $L$-QD.  There is an analogous virtual process  consisting of steps (1)-(3) in which the left TNW and $L$-QD are switched with the right TNW and $R$-QD, respectively. 
In the second process, an electron can be exchanged between TNWs without the double occupancy of QDs, \textit{e.g.} an electron tunnels from the $L$-QD to the left TNW, from the $R$-QD to the $L$-QD, and from the right TNW to the $R$-QD (and vice versa).

The first process is dominant if ${\epsilon_{0},\tilde{V}} \ll E_{R,L}^Z/2$  [see Fig.~\ref{Fig:EnDiag}(a)]. In addition, to ensure that one can neglect couplings to states above the superconducting gap, we require that $E_j^{\rm Z}/2\ll \Delta$. In this case, the coupling $D$ is given by
\begin{equation}\label{eq:D}
D=4\frac{t_M}{V}\frac{t_L t_R \left(E_L^\textrm{Z}+E_R^\textrm{Z}\right)}{E_L^\textrm{Z}E_R^\textrm{Z}}\,.
\end{equation}

The second process dominates in a regime similar to one described in Ref.~[\onlinecite{Silas1}], in which the QD levels are tuned out of the resonance with MBSs: ${E_j^\textrm{Z}} \ll \epsilon_0 \ll V, \Delta$ [see Fig.~\ref{Fig:EnDiag}(b)]. In this regime, the coupling $D$ has the following form:
\begin{equation}\label{eq:D2}
D=\frac{t_Mt_Lt_R\left(E_L^\textrm{Z}+E_R^\textrm{Z}\right)}{\epsilon_0^3}\,.
\end{equation}
The two terms in Eq.~(\ref{eq:D}) and Eq.~(\ref{eq:D2}) are of the same order of magnitude. However, we believe that the first regime, with ${\epsilon_0,\tilde{V} \ll E_j^\textrm{Z} /2\ll \Delta}$, is more optimal. The reason for this is twofold. (1) If ${|\epsilon_{0}|\gg E^{\rm Z}_{j}}/2$, the Zeeman splittings of the QDs are small, such that  $|T_+\rangle=d_{L,\uparrow}^{\dagger}d_{R,\uparrow}^{\dagger}|0_{\rm DQD}\rangle$ and $|T_-\rangle=d_{L,\downarrow}^{\dagger}d_{R,\downarrow}^{\dagger}|0_{\rm DQD}\rangle$ states become close in energy to the $|S\rangle$ and $|T_0\rangle$, enabling leakage out of the spin qubit subspace due to possibly spin-non-conserving tunnel matrix elements. (2) In order to tune the nanowire into the topological regime, the Zeeman energies $\Delta^{\rm Z}$ in the NWs need to be larger than the superconducting gap, $\Delta^{\rm Z}>\Delta$, while, in order to operate in the regime of Eq.~(\ref{eq:D2}), the Zeeman energies of the dots need to satisfy ${\Delta\gg|\epsilon_0|\gg E_j^\textrm{Z}}$. This implies that a significant difference of Zeeman energies (orders of magnitude) would need to be engineered between the QDs and the NWs when coupling the MBS qubit and the $ST_0$ qubit in the regime in which Eq.~(\ref{eq:D2}) is valid.  For these reasons, in the remainder of the manuscript, we focus on the first regime in which the $ST_0$-MBS coupling is given by Eq.~(\ref{eq:D}).

\section{Quantum Gates}\label{sec:Universality}

\begin{figure}[t!]
	\includegraphics[width=0.48\textwidth]{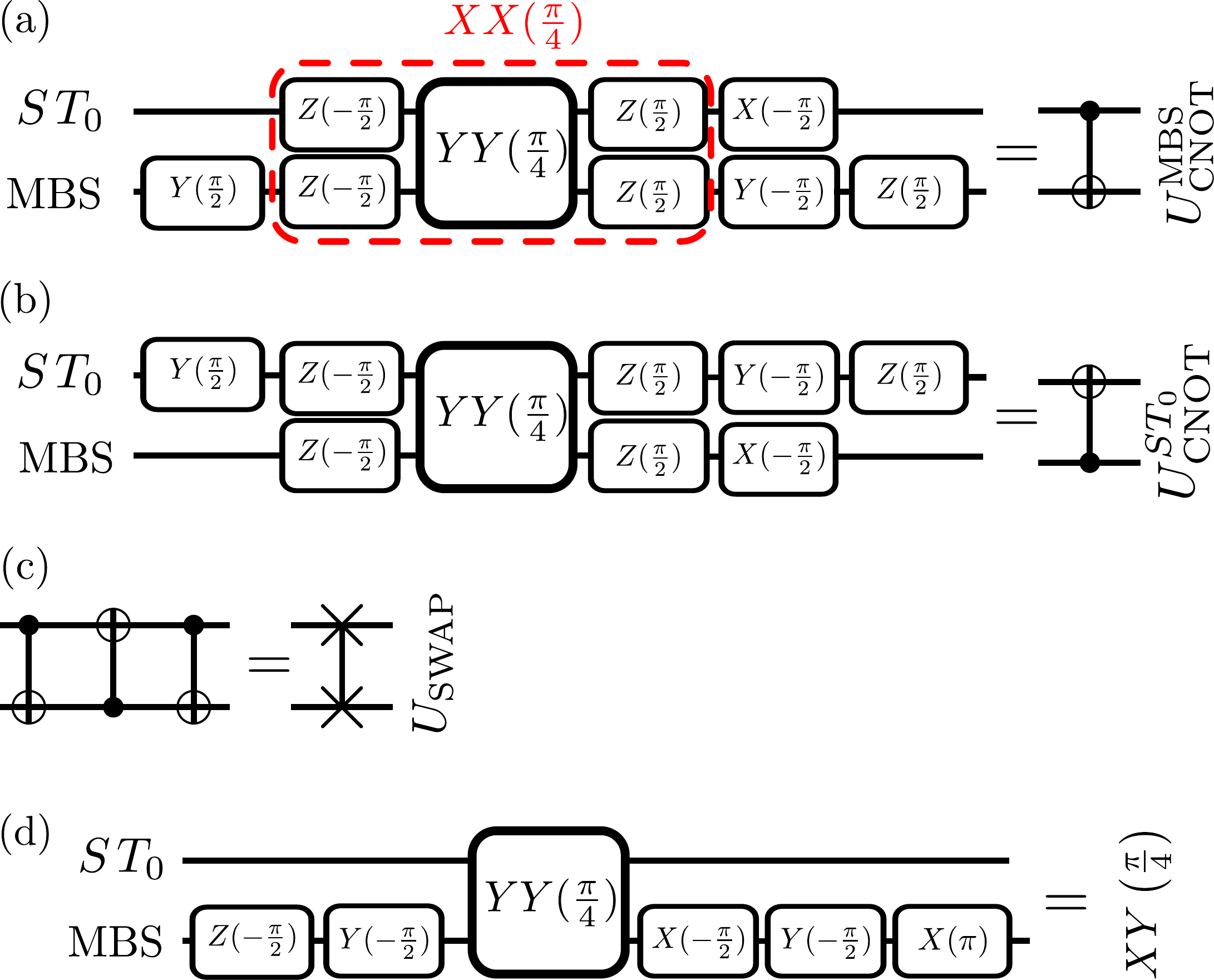}
	\caption{(Color online)
		 The protocol to generate additional two-qubit gates utilizing the $YY(\pi/4)$ gate and single qubit gates.  The subscripts $S$ and $M$ referring to a gate acting on the $ST_0$ and MBS qubits in the main text are omitted in this figure for simplicity. Operations by $X(\phi)$, $Y(\phi)$, or $Z(\phi)$ on the $ST_0$ or MBS qubit branch corresponds to a rotation about the $x$, $y$, or $z$ axis of the Bloch sphere, respectively, of the appropriate qubit. (a) An Ising $XX(\pi/4)$ gate can be obtained from an Ising $YY(\pi/4)$ gate using single qubit operations which can be further manipulated into a $U_\textrm {CNOT}^\textrm{MBS}$ gate. (b) The Ising $YY(\pi/4)$ gate is transformed into a $U_\textrm{CNOT}^{ST_0}$ gate by similar single qubit operations. (c) A $U_{SWAP}$ gate  can be obtained by successive application of three CNOT gates. (d) An Ising $XY(\pi/4)$ can be made from the Ising $YY(\pi/4)$ gate by  involving an appropriate exchange of MBSs.}\label{Fig2}
\end{figure}

After deriving the effective Hamiltonian $\mathcal{H}$ [see Eq.~(\ref{eq:LowEH})]  describing coupling between the MBS qubit and the $ST_0$ qubit, we explore which quantum gates can be generated with it.
Isolating the QDs from each other and from the MBSs, ${t_M=t_L=t_R=0}$, we reduce $\mathcal{H}$  to 

\begin{equation}
\mathcal H_X=\frac{E_L^\textrm{Z}-E_R^\textrm{Z}}{2}\sigma_x\,.
\label{H_X}
\end{equation}

 Waiting a time interval, $\tau$, induces a rotation about the $x$ axis on the Bloch sphere of the $ST_0$ qubit, 
 \begin{equation}
 X_S(\phi)=\exp{\left(-i \frac{E_L^\textrm{Z}-E_R^\textrm{Z}}{2\hbar}\tau\sigma_x\right)}\,,
 \end{equation} 
 by an angle ${\phi=(E_L^\textrm{Z}-E_R^\textrm{Z})\tau/\hbar}$. As selective control of the differences in Zeeman fields is experimentally difficult, a finite rotation about the $x$ axis always persists. This is typically overcome by using system parameters so that $J,D\gg \delta$ such that the error introduced in the gates is small. Alternatively, one can search for more involved schemes taking such terms into account\cite{PhysRevB.86.085423}. We account for this native imperfection in the discussion of the fidelities of the gates in the following section (Sec.~\ref{sec:Decoherence}) but disregard it in this section. Leaving the DQD uncoupled from the MBS while changing the interdot tunneling, $t_M$, rotations about the $z$ axis of the Bloch sphere of the $ST_0$ qubit can be selectively turned on and off. Pulsing $t_M$ for a time, $\tau$, induces a rotation about the $z$ axis on the Bloch sphere, 
 \begin{equation}
 Z_S(\phi)=\exp{\left(-i\frac{J}{\hbar}\tau\sigma_z\right)}\,,
 \end{equation}
 by an angle $\phi=2J\tau/\hbar$. Operations of $X_S(\phi)$ and $Z_S(\phi)$ allow full access to the $ST_0$ Bloch sphere. Furthermore, rotations about the $y$ axis of the Bloch sphere can be made by a composition of single qubit gates, ${Y_S(\phi)=Z_S(\pi/2)X_S(\phi)Z_S(-\pi/2)}$.

Fixed angle single qubit rotations on the Bloch sphere of the MBS qubit can be made by braiding MBSs, utilizing a nontopological part of the NW 
(see Fig.~\ref{Fig1}). By adiabatically exchanging the position of $\gamma_L$ and $\gamma_L'$\cite{alicea2011non}, the state of the MBS qubit is rotated by an angle $\pi/2$ about the $z$ axis of the Bloch sphere
\begin{equation}
Z_M\left(\frac{\pi}{2}\right)=\exp\left(-i\frac{\pi}{4}\eta_z\right)\,.
\end{equation}
 A $\pi/2$ rotation about the $x$ axis, 
 
 \begin{equation}
X_M\left(\frac{\pi}{2}\right)=\exp\left(-i\frac{\pi}{4}\eta_x\right)\,,
 \end{equation}
 is made by exchanging $\gamma_L'$ and $\gamma_R$.

The entangling coupling between the qubits is operational, $D\neq0$, only when all three tunnelings are finite, $t_M, t_L, t_R>0$. Disregarding rotations about the $x$ axis due to finite $\delta$, the Hamiltonian reduces to 
\begin{equation}
\mathcal H_+=J\sigma_z + D\sigma_y\eta_y\,.
\end{equation} 
By appropriately operating the tunnelings, one can perform the sequence of operations $Z_S(\phi)e^{-i \mathcal H_+ \tau_{YY}/\hbar}Z_S(\phi)$. For $\sin\phi=-J/D$, assuming $J\leq D$, and for $\tau_{YY}$ satisfying
\begin{equation}
\sin\left(\frac{\tau_{YY}\sqrt{D^2+J^2}}{\hbar}\right)=\sqrt{\frac{1+J^2/D^2}{2}}\,,
\end{equation}
we obtain the universal Ising gate 
\begin{equation}
YY\left(\frac{\pi}{4}\right)=\exp{\left(-i\frac{\pi}{4}\sigma_y\eta_y\right)}=\frac{1}{\sqrt{2}}\left(\mathbb 1 -i \sigma_y\eta_y\right)\,.
\label{YYgate}
\end{equation}
 
\noindent Although this realization of a $YY(\pi/4)$ gate is easily generated, it is only possible when $D\geq J$. In the following section, we use a set a parameters consistent with this condition and hence operate the entangling gate in this way. However, in the case $D< J$, a Trotter expansion can be used to approximately obtain a $YY(\pi/4)$ gate. \footnote[1]{For a general set of parameters, the term proportional to $D$ can be isolated using the Trotter formula\cite{trotterPAMS59}. First note that, upon single qubit operations, $\mathcal H_-=Y_S(-\pi)\mathcal H_+Y_S(\pi)=-J\sigma_z + D\sigma_y\eta_y$. A general $YY(\phi)$ gate can be obtained by\cite{trotterPAMS59}
\begin{equation*}
U_{YY}(\tau)=e^{-2i D\sigma_y\eta_y \tau/\hbar} \approx \left(e^{-i \mathcal H_+ \tau/n\hbar}e^{-i \mathcal H_- \tau/n\hbar}\right)^n\,,
\end{equation*}
where $n$ is the number of subdivisions in the time, $\tau$, in which $t_M$ is nonzero. Although the left and right sides of the equation are identical as $n\rightarrow\infty$, we obtain an approximate equality up to an arbitrary precision by increasing $n$. For $\tau=(\pi/8D)\hbar$, we obtain the gate $YY(\pi/4)=(\mathbb 1 -i \sigma_y\eta_y)/\sqrt{2}$ which entangles the $ST_0$ qubit and MBS qubit.}

One can use the available single qubit rotations on the $ST_0$ and MBS qubits to transform the universal Ising $YY(\pi/4)$ gate into a universal Ising $XX(\pi/4)$ gate\cite{Loss1} [see Fig.~\ref{Fig2}(a)],  

\begin{equation}
XX\left(\frac{\pi}{4}\right)=\exp{\left(-i\frac{\pi}{4}\sigma_x\eta_x\right)}=\frac{1}{\sqrt{2}}(\mathbb 1 -i \sigma_x\eta_x),
\end{equation} 
or the universal Ising $XY\left(\pi/4\right)$ gate [see Fig.~\ref{Fig2}(d)]\,,
\begin{equation}
XY\left(\frac{\pi}{4}\right)=\exp{\left(-i\frac{\pi}{4}\sigma_x\eta_y\right)}=\frac{1}{\sqrt{2}}(\mathbb 1 -i \sigma_x\eta_y)\,.
\end{equation}

Additional single qubit rotations further transform the $XX(\pi/4)$ gate into a CNOT gate using the $ST_0$ as the control and MBS qubit as the target qubit [see Fig.~\ref{Fig2}(a)],
\begin{equation}
 U_\textrm{CNOT}^{ST_0}=\frac{1}{2}(\mathbb 1-\sigma_z+\eta_x+\sigma_z\eta_x)\,,
\end{equation} 
or vice versa, using the MBS qubit as the control qubit and the $ST_0$ qubit as the target qubit [see Fig.~\ref{Fig2}(b)],
\begin{equation}
U_\textrm{CNOT}^\textrm{MBS}=\frac{1}{2}(\mathbb 1+\sigma_x+\eta_z -\sigma_x\eta_z)\,.
\end{equation} 
Sequential application of the CNOT gates yields the SWAP gate, ${U_{\textrm{SWAP}}=U_\textrm{CNOT}^\textrm{MBS}U_\textrm{CNOT}^{ST_0}U_\textrm{CNOT}^\textrm{MBS}}$ [see Fig.~\ref{Fig2}(c)]. $U_{\textrm{SWAP}}$ enables coherent exchange of quantum information between the $ST_0$ qubit and the MBS qubit. Any of the missing gates on the MBS can be performed by exchanging the quantum information, \textit{i.e.} by applying $U_{\textrm{SWAP}}$, applying the missing gate to the $ST_0$ qubit, and finally swapping back to the MBS qubits.
 
To readout the MBS one could likewise swap the quantum information into the $ST_0$ qubit and read it from there. However, a less gate-intensive procedure, and therefore potentially less error-prone, would be to form a Bell state between the qubits. The specific protocol to read a MBS qubit in a general initial state $\alpha|L\rangle+\beta|R\rangle$ is (1) prepare the $ST_0$ qubit in the singlet state, (2) apply the Ising $XY(\pi/4)$ gate between the $ST_0$ and the MBS qubit so that they are in a Bell state, and (3) perform the operation $Y_M(\pi/2)X_M(\pi)=(\mathbb 1-i\eta_y)\eta_x/\sqrt{2}$, \textit{i.e.} a Hadamard gate, on the MBS qubit. The resultant, entangled state is

 \begin{equation}\label{eq:mes}
 \frac{\beta-\alpha}{2}\left(|S\rangle |L\rangle-|T_0\rangle |R \rangle\right)+ \frac{\beta+\alpha}{2}\left(|S\rangle |R\rangle+|T_0\rangle |L\rangle
 	\right)\,.
 \end{equation}
Upon measuring the spin qubit, the probability to find it in the triplet state $|T_0\rangle$ is $|\alpha|^2$, while the probability to find it in the 
singlet state $|S\rangle$  is $|\beta|^2$. One can thereby readout the original superposition of states of the MBS qubit.

\section{Gate Fidelity}\label{sec:Decoherence} 

Having shown that our setup is in principle capable of universal quantum computation, in this section, we calculate the degree to which a physically implemented gate matches the ideal gate. The fidelity of an imperfect gate, $\tilde U$, as compared to an ideal gate, $U$, is defined as\cite{Pedersen1} 
\begin{equation}
\mathcal{F}=\frac{1}{n(n+1)}\left[{\rm tr}(MM^\dagger)+|{\rm tr}(M)|^2\right],
\end{equation} 
where $n$ is the dimension of the matrix representation of the gate and $M=U^\dagger \tilde U$; $\mathcal F=1$ if and only if $\tilde U=U$.

We include imperfections in the gate from two sources: (1) systematic error due to the hybridization of the singlet and triplet states, $\delta\neq0$, and (2) random fluctuations in the parameters as a result of coupling to the environment. The former modifies the gates by tilting the axis of rotation of the $ST_0$ Bloch sphere towards $x$ axis. To account for the latter, we describe the statistical distribution of the parameters with a normalized Gaussian distribution with a fixed mean and a standard deviation taken from experiments. That is, if $U(\tau)$ is the exact gate operation on the system as a result of a time evolution of the Hamiltonian for a time $\tau$, the imperfect gate is defined as
\begin{equation}
\tilde{U}(\tau)=\int p(\textbf{s})U(\tau) d\textbf{s}\,, 
\end{equation}
where we have defined the vector in parameter space $\textbf{s}=(t_L,t_R,t_M,E_L^\textrm{Z},E_R^\textrm{Z},V)$, $p(\textbf{s})=\prod_i{p(s_i)}$, and $p(s_i)=e^{(s_i-\bar s_i)^2/2\sigma_{s_i}^2}/\sqrt{2\pi\sigma_{s_i}^2}$ is a normalized Gaussian with mean $\bar s_i$ and standard deviation $\sigma_{s_i}$. As operations on the MBS qubit are believed to be topologically protected, we assume, for simplicity, that single qubit gates operating on the MBS qubit are perfect.

It is known that the decoherence time of the $ST_0$ qubit due to fluctuations in the Zeeman field\cite{Merkulov1,Petta1,Petta2,Rancic1} is much longer than the decoherence time due to electrostatic fluctuations\cite{dovzhenko2011nonadiabatic}. The standard deviation is proportional to the inverse of the decoherence time, thus, the standard deviation in the Zeeman fields turns out to be much smaller than the standard deviation in the parameters controlled by an electric potential or field, $\sigma_{E^\textrm{Z}_R},\sigma_{E^\textrm{Z}_L}\ll\sigma_{t_R},\sigma_{t_L},\sigma_V$. As such, we consider decoherence from the Zeeman field only when we operate the $X_S(\phi)$ gate and safely neglect it for all other operations. Moreover, as the Hamiltonian is proportional only to the difference in the Zeeman fields, $\delta$, when operating the $X_S(\phi)$ gate [Eq.~(\ref{H_X})], it is sufficient to specify the standard deviation in the differences of the Zeeman field, $\sigma_{\delta}$. 

In calculating the fidelity, we assume that the electrostatic control of our tunneling parameters is faster than any of our time scales in our system so that $t_M,~t_L$, and $t_R$ can be immediately turned on or off according to the gate operations in the previous section. For our estimates, we use typical experimental parameters for DQDs in InAs and Ge/Si core/shell NWs,\cite{Burkard1,Bjork1,Schroer1,hu2007ge}: $\bar t_R=\bar t_L=\bar t_M=100~\textrm{$\mu$eV}$, $\bar V=1~\textrm{meV}$, as well as $\bar E_L^\textrm{Z}=300~\mu$eV and $\bar E_R^\textrm{Z}=301.5~\mu$eV. 
These values are consistent with the assumptions made to obtain the effective Hamiltonian [see Eq.~(\ref{eq:LowEH})].  For this choice  of parameters, we get $\bar D\approx \bar J (4\bar t_M/\bar E_L^\textrm{Z})> \bar J$, and
the $YY(\pi/4)$ gate [see Eq.~(\ref{YYgate})] is performed on a timescale of $100\text{ ps}$, which is much faster than any spin\cite{camenzind2018hyperfine} or charge relaxations\cite{ChargeRelaxation} in a double quantum dot.

The standard deviations in the parameters controlled by the electrostatic fields are assumed to be equal to each other and the same for InAs and Ge/Si core/shell NWs: $\sigma_{t_L}=\sigma_{t_R}=\sigma_{t_M}=\sigma_{V}=0.4~\mu\textrm{eV}$ \cite{Martins1,Reed1}. As MBSs are realized in $n$-doped InAs NWs while Ge/Si core/shell NWs support MBS in $p$-doped NWs\cite{PhysRevB.90.195421}, the decoherence due to fluctuations in the Zeeman field can be rather different. Utilizing experimental data, the standard deviation in 
InAs NWs is $\sigma_{\delta}=0.07~\mu\textrm{eV}$\cite{Petta1,PhysRevLett.88.186802,PhysRevB.72.125337, PhysRevB.90.245305}. In contrast, because holes in Ge/Si  NWs have a $p$-type wavefunction, they only interact with nuclear spins via the dipole-dipole interaction and the orbital hyperfine interaction, leading to a much longer coherence time for holes as compared to electrons\cite{Fischer1}. Furthermore, Ge has $7.8\%$ isotopes with a non-zero nuclear spin while Si has $4.7\%$ of isotopes with a non-zero nuclear spin. Thus we assume $\sigma_{\delta}=0$ for the case of a Si/Ge core/shell NW. As we see below, the latter result provides a significant increase in the fidelities of two qubit gates when the $ST_0$-MBS qubit is embedded in a Ge/Si NW.

The fidelities of the two qubit gates between the MBS and $ST_0$ qubits are summarized in Tab.~\ref{Fig3}. We find the most error-free gates are the $YY(\pi/4)$ and the $XY(\pi/4)$ gate that have fidelities of $\mathcal{F}=0.9985$. Single qubit gates, which are themselves error-prone on the $ST_0$ qubit, must be used to transform $YY(\pi/4)$ to $XX(\pi/4)$, such that the fidelity of the latter gate decreases to $\mathcal{F}=0.993$. Likewise, operations of additional single-qubit gates in InAs (Ge/Si) result in a further decrease of the fidelities for $U_\textrm{CNOT}^\textrm{MBS}$ and $U_\textrm{CNOT}^{ST_0}$ to $0.989$ ($0.993$) and $0.961$ ($0.991$), respectively. As the single-qubit gates acting on the $ST_0$ qubit are imperfect, while those acting on the MBS qubit are decoherence free, the fidelity of $U_\textrm{CNOT}^\textrm{MBS}$ is greater than $U_\textrm{CNOT}^{ST_0}$. Lastly, because the SWAP gate is a product of three imperfect CNOT gates, it has a fidelity of $0.946$ ($0.978$) in InAs (Ge/Si) NWs. As expected, our fidelities always have a higher value for Ge/Si core/shell compared to InAs NWs due to lower decoherence from nuclear spins in Ge/Si.

We find that the Bell states obtained using the $XY(\pi/4)$ gate procedure, as described in the previous section, can be created with rather high fidelity in both InAs and Ge/Si core shell NWs. Furthermore, due to the fact that the error rate of the SWAP gate operated in an InAs NW is $\sim 5\%$, it is rather far away from the error rate of $1.1\%$ necessary to be amenable to the surface code \cite{gottesman2010introduction}. On the other hand, as the error rate of the SWAP gate acting on our hybrid qubit embedded in Ge/Si core/shell NWs is $\sim 2\%$, Ge/Si core/shell NWs are much more realistic candidate as the basis for a network of hybrid qubits.

\begin{table}
	\begin{center}
		\begin{tabular}{c||c|c|c|c|c}
			Gate&$YY\!\left(\frac{\pi}{4}\right)\!,XY\!\left(\frac{\pi}{4}\right)\!$&$XX\!\left(\frac{\pi}{4}\right)$&$U_\textrm{CNOT}^\textrm{MBS}$&$U_\textrm{CNOT}^{ST_0}$&$U_{\rm SWAP}$\\
			\hline
			$\mathcal{F}\text{(InAs)}$ & 0.9985  & 0.993  &  0.989  & 0.961 & 0.946\\
			$\mathcal{F}\text{(Ge/Si)}$ & 0.9985  & 0.993  &  0.993  & 0.991 & 0.978\\
		\end{tabular}
	\end{center}	
	\caption{Fidelities of several two qubit gates calculated using parameters for InAs and Ge/Si core/shell NWs. For both wires, we take the average values of the parameters to be the same: $\bar t_R=\bar t_L=\bar t_M=100~\mu\textrm{eV}$, $\bar V=1~\textrm{meV}$, $\bar E_L^\textrm{Z}=300~\mu$eV, and $\bar E_R^\textrm{Z}=301.5~\mu$eV. The standard deviation of the tunnelings and Coulomb repulsion are the same in both types of wires, ${\sigma_{t_L}=\sigma_{t_R}=\sigma_{t_M}=\sigma_V=0.4\,\mu\text{eV}}$, while ${\sigma_{\delta}=0.07\,\mu\text{eV}}$ for InAs and $\sigma_{\delta}=0$ for Ge NWs. As a result, the fidelities for the latter are significantly higher.
}\label{Fig3}	
\end{table}

\begin{figure}[t!]
\includegraphics[width=0.45\textwidth]{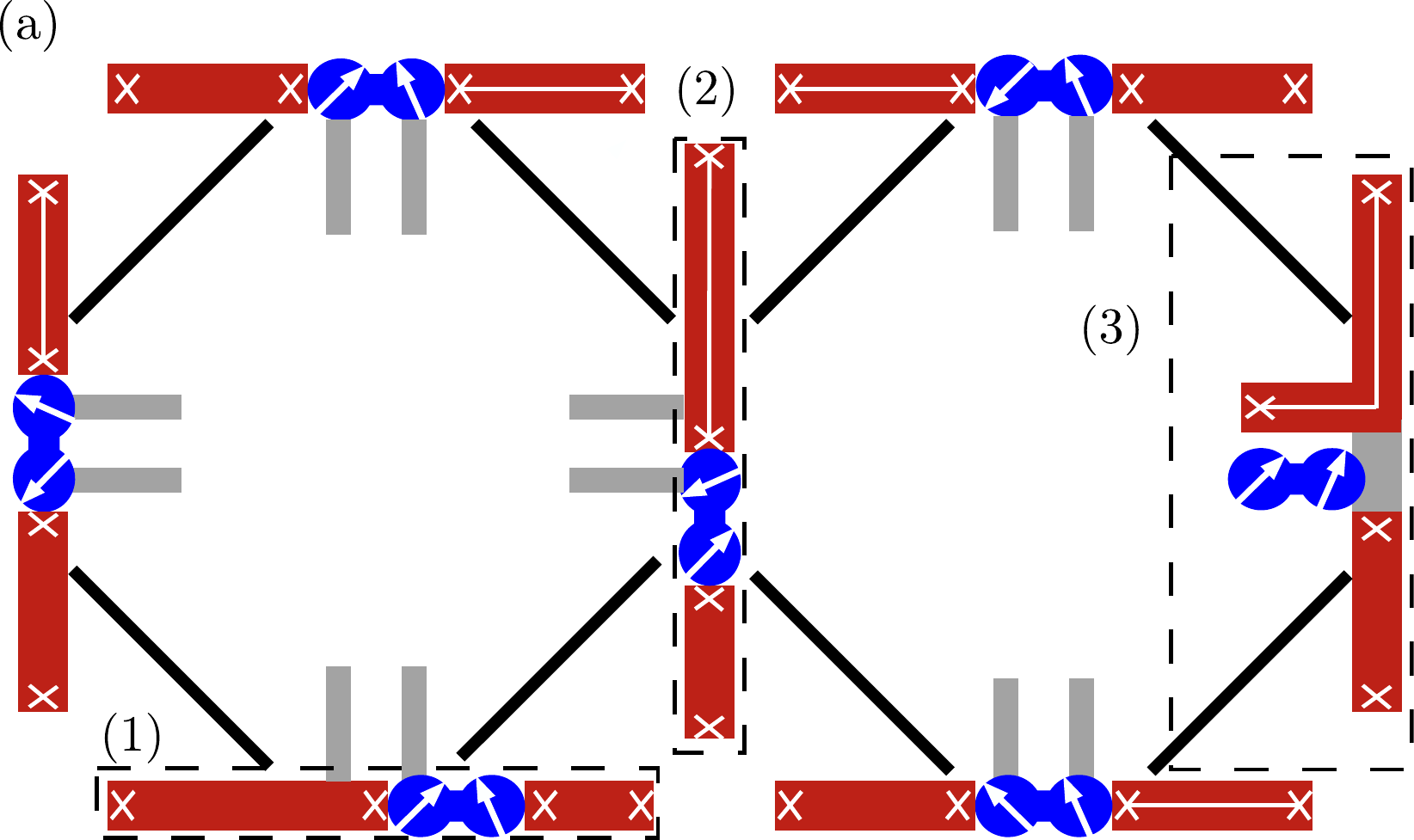}

\vspace{3 mm}
\includegraphics[width=0.45\textwidth]{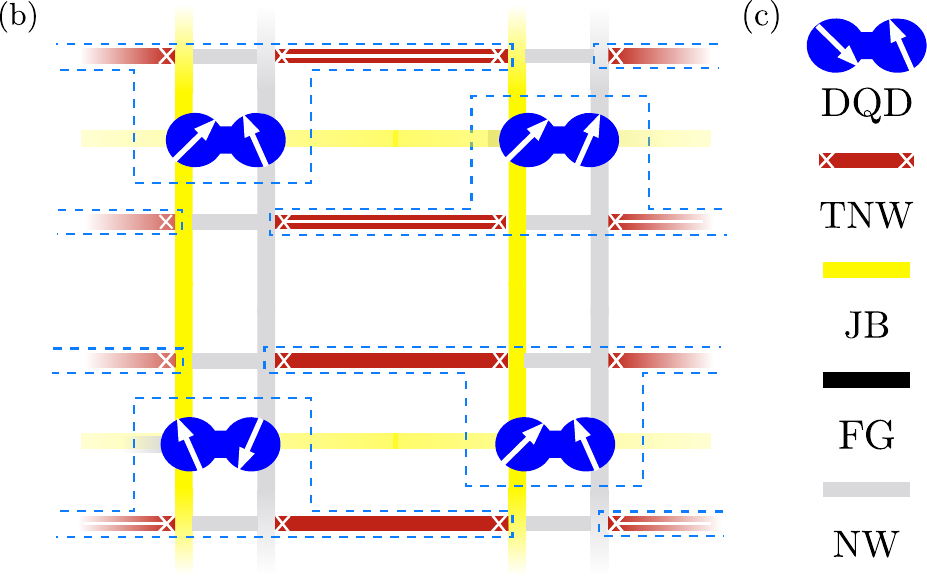}  
\caption{(Color online) (a) A network of $ST_0$-MBS hybrid qubits in the case when the TNWs and DQD are formed within the same NW [see Fig.~\ref{Fig1}(a)]. Qubits are selectively, long-range coupled by moving the DQDs in the neighborhood of a floating gate, \textit{e.g.} qubit (1) is coupled to qubit (2). Braiding of the MBSs can be performed by using the double T-junction as in qubit (3). (b) A network of $ST_0$-MBS hybrid qubits formed within a hashtag of NWs [see Fig.~\ref{Fig1}(b)]; a single hybrid qubit is drawn within the dashed blue line. Long-distance coupling between qubits is mediated by jelly bean quantum dots (JBs) that are tunably coupled to the DQDs. (c) The legend: DQD denotes a double quantum dot, TNW denotes a topological nanowire, JB denotes jelly bean quantum dot, FG denotes the floating gate and NW denotes a nanowire in the non-topological regime.}\label{Fig5}
\end{figure}

\section{Network of $ST_0$-MBS Qubit}\label{sec:Network} 
In this section we propose scalable 2D networks using $ST_0$-MBS qubits, with both the linear and hashtag setups. The basic unit in the case of the linear setup is the double T-junction [Fig.~\ref{Fig1}(a)]. Each double T-junction is connected to four adjacent qubits via a floating gate which terminates away from the DQDs [see Fig.~\ref{Fig5}(a)]. 
By moving two DQDs in adjacent qubits to the ends of the floating gate couples the $ST_0$ qubits, \textit{e.g.} in Fig.~\ref{Fig5}(a), qubits (1) and (2) are coupled. In this way a universal two-qubit CPHASE operation between the two coupled $ST_0$ qubits is performed, \textit{e.g.} ${U_\textrm{CP}=[\mathbb 1+\sigma_z^{(1)}+\sigma_z^{(2)}-\sigma_z^{(1)}\sigma_z^{(2)}]/2}$, where $\sigma^{(i)}_z$ is the Pauli matrix acting on the $i$th $ST_0$ qubit \cite{PhysRevB.75.085324,PhysRevX.2.011006,PhysRevB.95.245422}. The two auxiliary legs of the double T-junction are used to displace the DQD while the MBSs are braided [see Fig.~\ref{Fig5}(a)(3)].

In the second setup, the basic unit is the hashtag qubit [see Fig.~\ref{Fig1}(b)]. As it appears experimentally difficult to realize floating gates above NWs at present, we propose coupling a network of adjacent qubits, arranged in a hashtag geometry, in a manner which promises to be  experimentally more feasible [see Fig.~\ref{Fig5}(b)]. We propose to use the nontopological sections of the NWs as large quantum dots occupied with a large number of electrons and dubbed `jelly bean' quantum dots (JBs) in recent experiments\cite{baart2017coherent,malinowski2018fast} [see the yellow lines in Fig.~\ref{Fig5}(b)]. We assume that the tunneling between states on the QDs and the JB can be controlled by appropriate gating. Two adjacent hybrid qubits are uncoupled when the QD-JB tunneling is zero. Adjacent qubits are coupled by first decoupling the $ST_0$ qubit from the MBS qubit, $t_L=t_R=0$. Upon lowering the barrier between the QD and JB, an isotropic exchange interaction between one of the QDs and the JB is induced\cite{PhysRevB.96.201304}. Simultaneously coupling of two QDs in adjacent qubits to the JB induces an isotropic exchange interaction between the qubits, which can be utilized to perform universal two-qubit operations\cite{Loss1,Levy1,PhysRevB.96.201304}. This then provides a scalable network of pairwise connected qubits that allows one to emulate the surface code architecture.

As the MBS qubits are expected to be more robust against decoherence than spin qubits, we envision using the MBS qubits to store the information while the $ST_0$ qubits are used for processing. Alternatively, by both braiding the MBSs and coupling adjacent MBS qubits through the $ST_0$ qubit, one could perform operations directly on the MBS qubit. We also note that  by coupling topological wires to quantum dots, one can also test independently the presence of MBS \cite{pawel} going beyond the zero-bias feature that can also occur due to presence of trivial Andreev bound states \cite{abs1,abs2,abs3,reeg}.

\section{Conclusions}\label{sec:conclusion}
By coupling two topological superconductors hosting MBSs to a DQD, we have obtained an effective entangling coupling between an $ST_0$ qubit and a MBS qubit. Using this coupling, we show how to construct a Bell state between the two qubits that can be used to readout the MBS qubit via the $ST_0$ qubit. Furthermore, we realize a SWAP gate between the $ST_0$ qubit and the MBS qubit for which we estimate the fidelity to be $0.978$ by taking into account the native imperfections in the implementation of the gates and magnetic as well as electric noise typical in nanowire realizations of $ST_0$ qubits and MBSs. Lastly, we outlined a proposal to extend a single $ST_0$-MBS qubit to a scalable network. 

Although, we have found a scheme to realize universal quantum computation based  on MBS qubits with a fidelity close to the fidelity required to be amenable to the surface code\cite{gottesman2010introduction}, further increases of fidelities are possible. The fidelities are decreasing largely due to the noisy implementation of the single qubit gates acting on the $ST_0$ qubit and of the entangling gate. We believe this can be overcome by implementing dynamical decoupling sequences on the $ST_0$ qubit that is left for future work.

\acknowledgments
This work was supported by the Swiss National Science Foundation (Switzerland) and by the NCCR QSIT. This project received funding from the European Union's Horizon 2020 research and innovation program (ERC Starting Grant, grant agreement No 757725).

\bibliographystyle{apsrev4-1}
\bibliography{References4}
\clearpage

\end{document}